\documentclass{optica-article}

\journal{opticajournal} 

\articletype{Research Article}

\usepackage{lineno}

\begin{document}

\title{Absorption loss and Kerr nonlinearity in barium titanate waveguides}

\author{Annina Riedhauser,\authormark{1,\dag} Charles M$\ddot{\text{o}}$hl,\authormark{1,\dag}  Johannes Schading,\authormark{1,\dag} Daniele Caimi,\authormark{1}  David I. Indolese,\authormark{1}   Thomas M. Karg,\authormark{1} and Paul Seidler,\authormark{1,*}}

\address{\authormark{1}IBM Research Z$\ddot{\text{u}}$rich, Säumerstrasse 4, CH-8803 Rüschlikon, Switzerland

\authormark{\dag}The authors contributed equally to this work.}

\email{\authormark{*}pfs@zurich.ibm.com} 


\begin{abstract*} 
Because of its exceptionally large Pockels coefficient, barium titanate (BaTiO$_3$) is a promising material for various photonic applications at both room and cryogenic temperatures, including electro-optic modulation, frequency comb generation, and microwave-optical transduction. 
These applications rely on devices with low optical loss to achieve high efficiency.
Material absorption sets a lower limit to optical loss and is thus a crucial property to determine, particularly for integrated photonic devices.
Using cavity-enhanced photothermal spectroscopy, we measure the absorption loss of BaTiO$_3$ ridge waveguides at wavelengths near 1550~nm to be $\alpha_{\mathrm{abs}} = 10.9$~{\raisebox{0.5ex}{\tiny$^{+5.8}_{-0.4}$}} dB~m$^{-1}$, well below the propagation losses due to other sources, such as scattering.
We simultaneously determine that BaTiO$_3$ has a large Kerr nonlinear refractive index of $n_{\mathrm{2,BaTiO_3}}$ =  1.8 {\raisebox{0.5ex}{\tiny$^{+0.3}_{-0.3}$}} $\times$ 10$^{-18}$ m$^2$ W$^{-1}$. 
Considering these results, photonic integrated circuits utilizing BaTiO$_3$ have the potential to achieve significantly higher efficiency than demonstrated to date and are especially interesting for applications exploiting the combination of Pockels and Kerr effects.
\end{abstract*}

\section{Introduction}
Ferroelectric metal oxides are well known to exhibit large linear electro-optic (Pockels) coefficients \cite{bookAbel, SandoReview}
and have been used in bulk-crystal optical devices for decades \cite{Kaminow:66}.
Lithium niobate (LiNbO$_3$) in particular has become the workhorse for electro-optic modulators employed in optical communications \cite{KaminowApplPhysLett,Wooten,Zhang:21}.  With the advent of commercially available thin-film LiNbO$_3$ \cite{Levy, Rabiei, Poberaj}, a plethora of integrated photonic devices exploiting the Pockels effect have been developed \cite{Zhu:21}, including high-speed, broadband modulators \cite{li_lithium_2020}, frequency comb generators \cite{zhang_broadband_2019}, tunable lasers \cite{yin_electro-optically_2021,snigirev_ultrafast_2023}, and electro-optic quantum transducers \cite{holzgrafe_cavity_2020,mckenna_cryogenic_2020}.

More recently, BaTiO$_3$, a material with an exceptionally large Pockels coefficient ($r_{42} = 1640$~pm~V$^{-1}$ \cite{Johnston:65} compared to $r_{33} = 31$~pm~V$^{-1}$ for LiNbO$_3$ \cite{weis_lithium_1985}), has emerged as a promising alternative for integrated electro-optic devices because of the development of methods to grow high-quality epitaxial thin films on silicon \cite{abel_strong_2013}.
Abel \emph{et al.} \cite{abel_large_2019} have shown that the Pockels effect in thin-film  BaTiO$_3$ can be almost as strong as in bulk crystals, with $r_{42} = 923$~pm V$^{-1}$.
Thin-film BaTiO$_3$ electro-optic modulators were realized both with hybrid silicon waveguides and as plasmonic devices, achieving data rates up to 50~Gbit~s$^{-1}$. Multilevel non-volatile phase shifters have also been reported \cite{geler-kremer_ferroelectric_2022}. The Pockels coefficient remains as high as 200\,pm\,V$^{-1}$ at 4\,K, making BaTiO$_3$ a promising candidate for efficient low-temperature high-speed modulation \cite{eltes_integrated_2020}, as may be needed for various photonic quantum technologies \cite{psiquantum}. 
Thin-film BaTiO$_3$ thus has the potential to displace LiNbO$_3$ in a variety of future integrated photonic circuits.

In addition to a large Pockels coefficient, efficient modulation of light with electric signals requires low optical propagation loss in the nonlinear material \cite{puckett_422_2021}. Optical loss can have several origins, such as scattering,  absorption, insufficient guiding in bends, or coupling to other waveguide modes. Scattering can occur at surfaces or interfaces due to roughness \cite{lee_ultra-low-loss_2012,pfeiffer_ultra-smooth_2018} and within the material as a result of index variations at domain boundaries. Absorption loss is associated with bulk defects and impurities, such as traces of metals \cite{pfeiffer_ultra-smooth_2018}, hydrogen \cite{puckett_422_2021, liu_high-yield_2021}, and hydroxyl groups \cite{heinemeyer_annihilation_2006, churaev_heterogeneously_2023, wu_oh_2023}, as well as surface species \cite{parrain_origin_2015} and even adsorbed water \cite{gorodetsky_ultimate_1996}. 
Absorption also occurs due to single- and multi-photon electronic transitions in the pure material \cite{husko_multi-photon_2013}.  While some optical loss channels can be mitigated by engineering, such as scattering due to surface roughness or bending loss, the material's inherent absorption sets a lower achievable limit.

In the case of the more-established thin-film LiNbO$_3$ platform, considerable progress has already been made in reducing the various sources of loss.  
Ridge waveguides made from ion-sliced thin-film LiNbO$_3$ defined by dry etching have achieved propagation loss of $2.7$~dB~m$^{-1}$ \cite{Zhang:17}. 
Moreover, it has been determined that the material absorption-limited propagation loss in these waveguides, if annealed in an oxygen atmosphere, can be as low as $0.2 $~dB~m$^{-1}$ \cite{Shams-Ansari:1} and that other loss sources such as scattering dominate.

In contrast, work on loss reduction is less advanced for waveguides made of thin-film BaTiO$_3$.
The extremely high propagation loss initially observed in various hybrid BaTiO$_3$ waveguides was found to be caused by absorption associated with hydrogen incorporated in the thin strontium titanate seed layer typically used for the epitaxial deposition of BaTiO$_3$ on silicon \cite{eltes_low-loss_2016}.  Depending on the specific waveguide structure, propagation losses could be reduced to 200 to 600 dB~m$^{-1}$ by annealing under conditions that drive out the hydrogen and simultaneously avoid the formation of oxygen vacancies.
More recently, losses as low as 70 dB~m$^{-1}$ were achieved in heterogeneously integrated BaTiO$_3$-Si$_3$N$_4$ hybrid waveguides \cite{Riedhauser:1}.  In BaTiO$_3$-only ridge waveguides, for which more of the optical mode is confined in the BaTiO$_3$, losses below 200~dB~m$^{-1}$ \cite{Mohl:22} have been achieved, but this is still two orders of magnitude higher than the propagation losses of similar waveguides made of LiNbO$_3$, and it is an open question if this is due to an inherent material limitation.  
Here, we determine the material absorption loss near 1550 nm of thin-film BaTiO$_3$ ridge waveguides using a modulation transfer technique based on cavity-enhanced photothermal spectroscopy \cite{rokhsari_observation_2005, Gao:1, Shams-Ansari:1}.  This method has proven useful for many other photonic platforms \cite{liu_high-yield_2021, Wilson:1} and allows us to simultaneously determine the Kerr nonlinearity of BaTiO$_3$.  We measure a median absorption-limited propagation loss in BaTiO$_3$ of $10.9$ {\raisebox{0.5ex}{\tiny$^{+5.8}_{-0.4}$}}~dB~m$^{-1}$, 
suggesting that it should be possible to produce resonators with intrinsic quality
factors well above $10^6$.  
The Kerr nonlinear refractive index $n_{\mathrm{2,BaTiO_3}} = 1.8$ {\raisebox{0.5ex}{\tiny$^{+0.3}_{-0.3}$}} $\times$  $10^{-18}$~m$^2$~W$^{-1}$ is an order of magnitude higher than that of both LiNbO$_3$ and the widely used nonlinear optical material Si$_3$N$_4$, 
opening numerous opportunities for active photonic integrated circuits based on BaTiO$_3$.

\section{Device description and characterization}

\begin{figure}[htbp]
\centering\includegraphics[width=7cm]{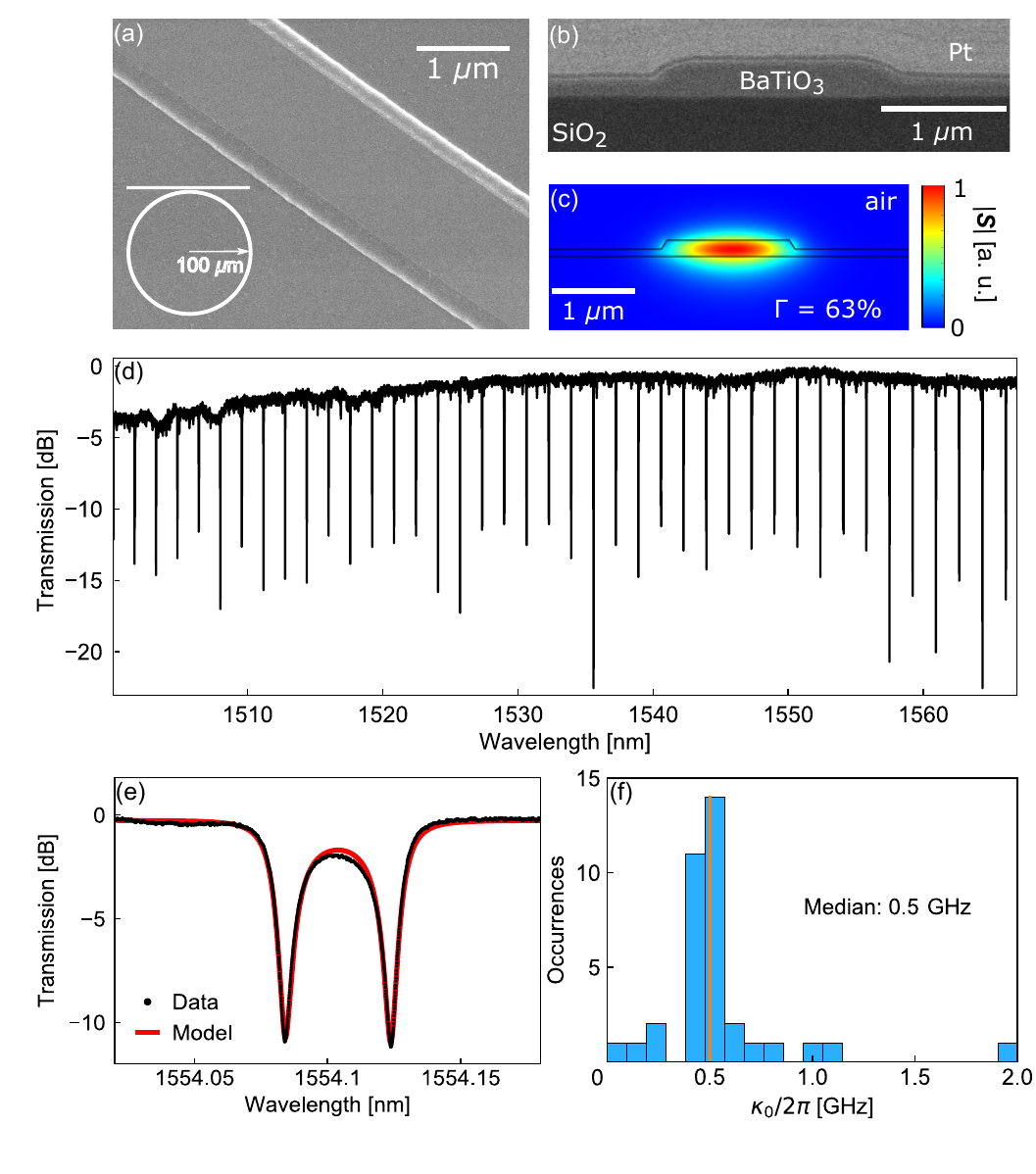}
\caption{(a) Scanning electron microscope (SEM) image of a BaTiO$_3$-on-silicon-dioxide ridge waveguide forming a ring resonator with a radius of 100 $\mu$m. (b) SEM image of a waveguide cross-section prepared by focused ion beam (FIB) milling. The waveguide is coated with platinum (Pt) solely for preparation of the cross-section. (c) Finite-element-method simulation of the Poynting vector magnitude |$\textbf{S}$| of the TE$_{00}$ waveguide mode (linear color scale).
(d) Measured transmission spectrum of a ring resonator with a radius of 100~$\mu$m. (e) Split resonance at 1554 nm and fit to two Lorentzians. (f) Distribution of intrinsic loss rates $\kappa_0$ of the resonances of a ring resonator. The orange vertical line indicates the median.}
\label{fig:SEM_spectra}
\end{figure}

The modulation transfer experiment is performed on air-clad BaTiO$_3$-on-silicon-dioxide ridge-waveguide microring resonators with a radius of 100 $\mu$m [Fig. \ref{fig:SEM_spectra}(a)]. The fabrication process is similar to that published in prior work \cite{abel_large_2019, eltes_integrated_2020, Mohl:22}. A multi-domain crystalline BaTiO$_3$ film with a thickness of 225~nm grown epitaxially on a single-crystalline silicon substrate is bonded to a silicon wafer capped with 3~$\mu$m of thermal silicon dioxide followed by removal of the silicon substrate on which the BaTiO$_3$ was grown.
The ring resonators are defined by electron-beam lithography and the pattern is transferred to BaTiO$_3$ by inductively coupled-plasma reactive-ion etching.  The sample is finally annealed for several hours at 600$^{\circ}$C in an oxygen atmosphere, to fill potential oxygen vacancies, remove incorporated hydrogen \cite{eltes_low-loss_2016}, and heal any damage from electron beam lithography \cite{shi_reduced_2024}.

The waveguides have a ridge height of 110~nm
and a width of 1.6~$\mu$m at the top of the ridge [Fig. \ref{fig:SEM_spectra}(a),(b)].  
A finite-element-method simulation of the fundamental transverse-electric TE$_{00}$ mode of the ridge waveguide [Fig. \ref{fig:SEM_spectra}(c)] yields an optical mode confinement in BaTiO$_3$ of $\Gamma = \iint_{BaTiO_3} |\textbf{S}|dA/ \iint |\textbf{S}| dA = $ 63\%, where $\textbf{S}$ is the Poynting vector. The surface integral in the numerator is over the area covered by BaTiO$_3$, whereas the surface integral in the denominator is over the entire cross-section. 

We initially characterize the ring resonators by measuring the transmission spectrum for transverse-electric polarized light [Fig. \ref{fig:SEM_spectra}(d)]. 
Most resonances are split due to scattering at imperfections that couples 
clockwise- and counterclockwise-propagating modes \cite{puckett_422_2021}.  We fit the split resonances with a model describing coupled Lorentzians [Fig. \ref{fig:SEM_spectra}(e)] to extract the  intrinsic loss and external coupling rates, $\kappa_0$ and $\kappa_{ex}$, respectively (see Supplement 1 for details).  A typical distribution of the intrinsic loss rates for one device is shown in Fig. \ref{fig:SEM_spectra}(f).
The median value of $\kappa_0/2\pi$ for each of the eight devices measured ranges between 430~MHz and 870~MHz and averages 510~MHz (see Supplement 2 for the loss distributions of all eight devices). The corresponding range of intrinsic quality factors $Q_0 = \omega/\kappa_0$, where $\omega$ is the angular resonance frequency, extends from 2.22 $\times$ 10$^5$ to 4.50 $\times$ 10$^5$, corresponding to propagation losses between 86~dB~m$^{-1}$ and 174~dB~m$^{-1}$.\\
\begin{figure}[htbp]
\centering\includegraphics[width=14cm]{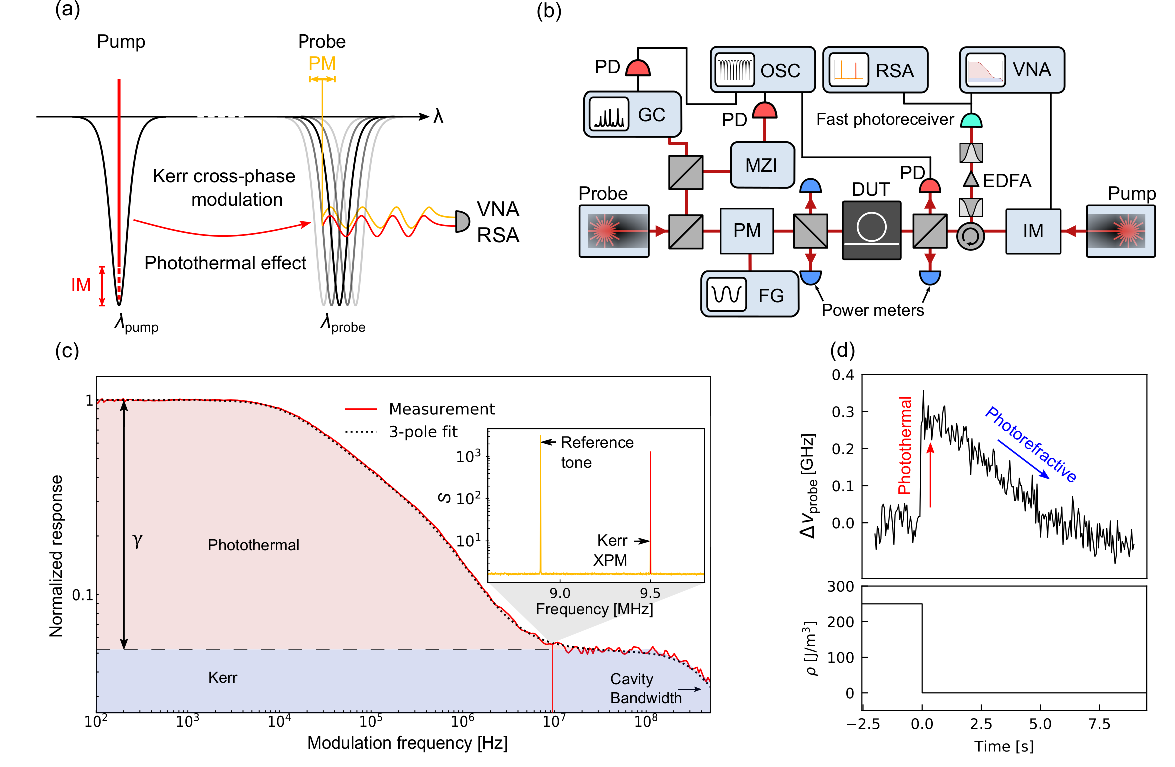}
\caption{(a) Illustration of the modulation transfer experiment. An intensity-modulated laser pumps the ring resonator on resonance while another phase-modulated laser probes the photothermal and Kerr response on the flank of another resonance. (b) Schematic of the experimental apparatus used to measure the response curve and the photorefractive lifetime. After coming out of the device under test (DUT), the probe light is first filtered to remove the reflected pump light before being amplified and detected on a fast photoreceiver with 500 MHz bandwidth and finally sent to either a VNA (response curve) or RSA (calibration). The transmission spectra measured on the oscilloscope (OSC) are referenced to the laser frequency measured in parallel using a Mach-Zehnder interferometer (MZI) and to a gas cell (GC) for absolute wavelength calibration. PM phase modulator, FG function generator, IM intensity modulator, EDFA erbium-doped fiber amplifier, PD photodetector, RSA real-time spectrum analyzer, VNA vector network analyser. (c) Measured modulation transfer response curve with 3-pole fit (dashed line)
used to extract $\gamma$. Inset: Probe-laser power spectral density $S(\nu)$ showing the phase-modulated reference and cross-phase-modulated (XPM) signals. (d) Photorefractive lifetime measurement: (top) Dynamics of the probe resonance frequency shift after switching off the pump laser. (bottom) Energy density in resonator as a function of time.}
\label{fig:modulation_transfer}
\end{figure}

\section{Measurement of absorption loss and Kerr nonlinearity}
\label{Sec:MTR}

Next, we use a modulation-transfer technique to determine the photothermal and Kerr nonlinearities \cite{rokhsari_observation_2005, Wilson:1, Gao:1, Shams-Ansari:1} of the BaTiO$_3$ waveguides. As illustrated schematically in Fig. \ref{fig:modulation_transfer}(a), the ring resonator is pumped with intensity-modulated laser light on resonance with one of its modes, while a second, weaker laser field is tuned to the flank of a different resonance and probes its frequency modulation in response to the pump.
Due to the distinct timescales of the photothermal and Kerr response, the two effects can be distinguished by recording the signal as a function of modulation frequency.  In our experimental apparatus [Fig. \ref{fig:modulation_transfer}(b)], light from the pump and probe lasers is injected into the device in opposite directions, so that only reflected pump light needs to be filtered out of the probe signal.  The amplified probe signal is detected with a fast photoreceiver. The frequency-dependent response curve [(Fig. \ref{fig:modulation_transfer}(c)] is obtained using a vector network analyzer to drive the intensity modulator and record the probe signal. (Typical experimental parameters are provided in Supplement 3.)
The observed signal exhibits two characteristic plateaus as a function of modulation frequency. At frequencies below 1~MHz, the response is dominated by the photothermal effect associated with optical absorption. The second plateau emerging above 10 MHz is attributed to the Kerr effect. 
A final frequency cut-off is determined by the total cavity dissipation rate $\kappa = \kappa_0 + \kappa_{ex}$.

Note that we observe no dependence of $\kappa_0$ on the low circulating powers employed, excluding significant loss contributions from nonlinear effects such as two-photon absorption or second-harmonic generation. This is consistent with the low cutoff wavelength of BaTiO$_3$ (electronic bandgap $\approx$ 3.2 eV,\cite{ tihtih_structural_2022,habdulhadi_conserved_2022}) and the fact that there should be no resonant enhancement of second-harmonic generation in our devices, as dispersion makes it unlikely that there will be a phase-matched resonance at the second-harmonic frequency.

To determine the absorption loss rate, the response curve must be calibrated.  A typical method uses the dependence of the resonance frequency shift on pump laser power \cite{Gao:1,Wilson:1}, which measures the sum of Kerr and photothermal effects.  Combined with the ratio of the photothermal and Kerr plateau levels and knowledge of the thermorefractive coefficient as well as the rate of temperature change with absorbed optical power, the absolute value of both effects can be determined.  This quasi-static measurement assumes however that there are no other physical effects occurring at low or zero modulation frequency.  In LiNbO$_3$, for example, the photorefractive effect acts on a timescale of tens of milliseconds to minutes \cite{Shams-Ansari:1, Jiang:17, Xu:21}, so the level of the photothermal plateau cannot be reliably determined in a quasi-static measurement.  Pyroelectricity may also interfere with measurement of the photothermal response at slow timescales \cite{zhang_pyro}. To ascertain whether such effects play a role in our platform, we perform an experiment in which we monitor the time evolution of the probe resonance frequency after switching off the pump laser (Fig. \ref{fig:modulation_transfer}(d)). We observe an initial rapid increase in the probe resonance frequency due to interruption of the photothermal and Kerr effects, followed by a slow decrease over a timescale of several seconds that we attribute to the photorefractive effect \cite{Shams-Ansari:1}. Although we cannot definitively rule out a pyroelectric effect, we have not seen any evidence of it in experiments where we change the temperature of the sample and monitor the shift of optical resonances.  In a static optical pump-probe experiment carried out at low power levels to minimize the photothermal effect, we find that the probe resonance blue shifts with increasing pump power, consistent with the photorefractive effect (Supplement 4).

Based on these observations, we choose to calibrate the modulation-transfer response by comparing the Kerr cross-phase modulation amplitude to a known phase modulation of the probe laser \cite{Shams-Ansari:1, Gorodetksy:10} at a frequency near the Kerr plateau, well beyond the timescale of the photorefractive effect. Specifically, we simultaneously phase modulate the probe laser at a frequency of $\nu_{\mathrm{PM}} = 8.9$~MHz and intensity modulate the pump light at a slightly different frequency $\nu_{\mathrm{IM}} = 9.5$~MHz. Modulation depths of the phase ($\beta_{\mathrm{PM}}$) and intensity ($\beta_{\mathrm{IM}}$) modulators were calibrated using balanced heterodyne detection. With a real-time spectrum analyzer, we measure the power spectral density of the probe laser intensity noise and use the phase-modulation reference peak to calibrate the Kerr cross-phase modulation peak [Fig.\ref{fig:modulation_transfer}(c) inset], from which we
determine the Kerr-induced cross-phase cavity frequency shift $\delta \nu_{\mathrm{XPM}}$ (see Supplement 5 for further details).
The effective nonlinear index $\overline{n_2}$ of the waveguide mode can then be extracted from the cross-phase modulaton signal and the calibrated intensity-modulation depth $\beta_{IM}$ using (\cite{Shams-Ansari:1} and Supplement 5): 
\begin{equation}
\overline{n_2} = \dfrac{\delta \nu_{\mathrm{XPM}} \overline{n_0n_g}}{2 c \nu \beta_{\mathrm{IM}} \rho},
\label{n2}
\end{equation}
where $\rho$ is the energy density of the pump light in the resonator as determined from the measured values of $\kappa_0$ and $\kappa_{ex}$, $c$ is the speed of light, $\nu$ is the frequency of the unperturbed probe resonance, $n_0$ is the material refractive index, and $n_g$ is the group index considering only material dispersion.  The bar indicates that the value of the respective variable is the average weighted by the optical mode intensity distribution (see Supplement 5).

The median value of $\overline{n_2}$ calculated from the calibrated response curves of the eight nominally identical devices measured is 1.7 {\raisebox{0.5ex}{\tiny$^{+0.3}_{-0.3}$}} $ \times$ 10$^{-18}$~m$^2$W$^{-1}$. The indicated errors are the distances of the median to the 25\% (-) and 75\% (+) quantiles. From this value and the known value of $n_2$ for SiO$_2$ \cite{Flom:15}, we determine the nonlinear index of BaTiO$_3$ to be $n_{2,\mathrm{BaTiO_3}} = 1.8$ {\raisebox{0.5ex}{\tiny$^{+0.3}_{-0.3}$}} $\times 10^{-18}$~m$^2$~W$^{-1}$ (Supplement 6).
Using the ratio $\gamma$ of the levels of the photothermal and Kerr response curve plateaus, we compute the absorption rate
\begin{equation}
\kappa_{\mathrm{abs}} = \dfrac{-\delta\nu_\mathrm{XPM} \gamma}{\beta_{\mathrm{IM}}\rho V_{\mathrm{eff}} \dfrac{\mathrm{d}\nu}{\mathrm{d}T} \cdot \dfrac{\mathrm{d}T}{\mathrm{d}P_{\mathrm{abs}}}}.
\label{kappa_abs}
\end{equation}
The change in frequency with temperature due to the thermorefractive effect, $\dfrac{\mathrm{d}\nu}{\mathrm{d}T}$, is measured by changing the temperature of the chip and monitoring the resonance frequency shift.  The rate of change of temperature with absorbed power $\dfrac{\mathrm{d}T}{\mathrm{d}P_{\mathrm{abs}}}$ is estimated from a finite-element simulation (Supplement 7), as is the effective volume $V_{\mathrm{eff}}$ of the mode (Supplement 1). We obtain a median value for $\kappa_{\mathrm{abs}}/2\pi $ of 55 {\raisebox{0.5ex}{\tiny$^{+29}_{-2}$}} MHz. (Values of $\kappa_{\mathrm{abs}}/2\pi$ and $n_{2,\mathrm{BaTiO_3}}$ for all the measured devices are given in Supplement 8.) This corresponds to an absorption-limited propagation loss of $\alpha_{\mathrm{abs}} = 10.9$ {\raisebox{0.5ex}{\tiny$^{+5.8}_{-0.4}$}} dB~m$^{-1}$. As the underlying thermal silicon oxide has negligible absorption, we conclude that the measured absorption loss is associated with the BaTiO$_3$. The total propagation loss rates of our BaTiO$_3$ ridge waveguides are roughly an order of magnitude larger than the absorption rate, suggesting that other loss sources such as scattering are dominant in our devices. \\

\section{Discussion and outlook}
In conclusion, we have determined the material absorption-limited propagation loss near 1550\,nm in thin-film BaTiO$_3$ ridge waveguides, as well as the Kerr nonlinear index, using cavity-enhanced photothermal and Kerr cross-phase modulation spectroscopy.  Table \ref{tab:comparison} compares the values obtained with those of other state-of-the-art nonlinear materials. While the absorption loss of the sample of BaTiO$_3$ we measured is still more than an order of magnitude larger than achieved in LiNbO$_3$, it should be noted that the growth and processing of BaTiO$_3$ thin-films is being continually improved.  Given the relatively short development history of BaTiO$_3$ as a material for integrated photonics, it is reasonable to expect further reduction in optical absorption.
Possible avenues are more sophisticated anealing \cite{shi_reduced_2024} and lithography methods \cite{zhu_twenty-nine_2024}, similar to the development of low-loss LiNbO$_3$ waveguides. An important question for future optimization is the origin of the remaining loss channels apart from absorption, which could for example be scattering within the material or at the surfaces. Prior work on the BaTiO$_3$ platform has shown a reduction in propagation losses with increasing waveguide width, suggesting that sidewall roughness is a significant loss channel \cite{Mohl:22}.
It is thus likely that the achievable efficiency of integrated electro-optic devices utilizing BaTiO$_3$ will continute to improve. 
Because its Pockels effect remains strong at cryogenic temperatures \cite{eltes_integrated_2020}, BaTiO$_3$ is also a promising candidate for low-temperature applications requiring high sensitivity, such as focal plane arrays \cite{johnston}, superconducting-nanowire single-photon detectors \cite{de_cea_photonic_2020}, cryogenic modulators \cite{youssefi_cryogenic_2021, shen_traveling-wave_2024}, and microwave-optical quantum transducers for superconducting qubit control and readout \cite{arnold2023, warner2023}.

We also found that our
BaTiO$_3$ waveguides exhibit a significant Kerr nonlinearity,
about an order of magnitude larger than that of LiNbO$_3$ as well as that of Si$_3$N$_4$, the most commonly used material for high-performance photonic integrated circuits exploiting third-order optical nonlinearity \cite{Pfeiffer:17}. 
There are few previous studies quantifying the Kerr nonlinearity in BaTiO$_3$ \cite{zhang_nonlinear_2000, Ganeev:08, yust_enhancement_2012}, and the published values range over six orders of magnitude.  One reported a value of $n_2 = 6 \times 10^{-20}$~m$^2$~W$^{-1}$ for a 2~mm-thick bulk crystalline sample measured using the z-scan technique with a pulsed light source at 790~nm but a value nearly four orders of magnitude larger for nanoparticles in a suspension \cite{Ganeev:08}. Another reported values of $n_2$ in the range of  $10^{-14}$~m$^2$~W$^{-1}$ for nanoparticles, again measured with the z-scan method but with a continuous-wave light source at 532~nm \cite{yust_enhancement_2012}.
The differences in probe wavelength, measurement technique, material synthesis, and sample dimensions preclude a simple comparison with our results.  (See Supplement 9 for further discussion.) 
Nevertheless, the presence of the Kerr nonlinearity in BaTiO$_3$ provides an opportunity for additional functionality.
As can be seen from Table 1, most Kerr nonlinear materials, including III-V compound semiconductors, for which $n_2$ can be quite large, have small or zero Pockels coefficients.  Ferroelectrics and especially BaTiO$_3$ are thus attractive choices for monolithically integrated devices requiring both a large Pockels effect and a large Kerr nonlinearity \cite{wang_monolithic_2019}.  Such devices could be used for high-precision optical clocks \cite{Papp:14}, soliton microcomb ranging with adjustable update time and range ambiguity \cite{doi:10.1126/science.aao1968}, parallelized data communication using solitons to achieve high data rates over large bandwidth \cite{marin-palomo_microresonator-based_2017}, or high-resolution dual-comb spectroscopy \cite{doi:10.1126/science.aah6516}.

\begin{table}[htbp]
	\centering
	\begin{tabular}{cccc}
		\toprule
    		Material & $n_{\mathrm{2}}$ (10$^{-20}$ m$^2$ W$^{-1}$) & $r$ (pm V$^{-1}$) & ${Q_{\mathrm{abs}}}$(10$^6$) \\
	     	\midrule
		SiO$_2$ 	& 2.2  & 0 & 3900 \\
		Si$_3$N$_4$ & 24  & 0 & 290 \\
		LiNbO$_3$ 	& 17  & 30 & 163 \\
        BaTiO$_3$  & 180 (this work) & 923 \cite{abel_large_2019} & 3.8 (this work) \\
		GaP 					& 1130  & 1.1 @ 1153 nm\cite{nelson_electrooptic_1968} & - \\
		Al$_{0.2}$Ga$_{0.8}$As 		& 2600  & 1.5 @ 1520 nm \cite{berseth_electrooptic_1992} & 2.0 \\
		
	    \bottomrule
	\end{tabular}
	\caption{Comparison of the Kerr nonlinear index $n_2$, the Pockels coefficient $r$, and the material absorption-limited quality factor $Q_{\mathrm{abs}} = \omega/\kappa_{\mathrm{abs}}$ near 1550 nm for several state-of-the-art integrated photonic platforms. Unless stated otherwise, the values are taken from \cite{Gao:1} for SiO$_2$, Si$_3$N$_4$ and Al$_{0.2}$Ga$_{0.8}$As, from\cite{Shams-Ansari:1} for LiNbO$_3$, and from \cite{Wilson:1} for GaP.}	
	\label{tab:comparison}
\end{table}

\newpage

\begin{backmatter}
	
\bmsection{Supplementary Material}
The supplementary material includes a derivation of the analytical expression used
to fit the resonances in the transmission spectra and the expressions used to compute
the circulating energy density in the resonators. We also provide the distribution of intrinsic loss rates for all measured devices, the measurement procedure and parameters employed, and additional evidence for the existence of a photorefractive effect.  We describe the method for calibration of the Kerr cross-phase frequency modulation, 
the extraction of the Kerr nonlinear index of BaTiO$_3$, the determination of the thermorefractive shift, and the simulation of the thermal response due to absorption of light.  Lastly, we give the values of the Kerr nonlinearity and absorption losses obtained for all measured devices.

\bmsection{Author Contributions}
A.R. designed the sample. C.M. developed the processing steps for fabrication of BaTiO$_3$ ridge waveguides. D.C. fabricated the sample with help from A.R.. A.R. and J.S. performed the characterization measurements and data analysis with guidance from D.I.I. and T.M.K.. A.R. carried out the thermal response simulations. A.R., T.M.K., D.I.I, C.M., and P.S. wrote the manuscript. P.S. conceived and supervised the project.

\bmsection{Funding}
This work was supported by the Swiss National Science Foundation under grant agreement No. 186364 (QuantEOM) and by the European Union Horizon 2020 Research and Innovation Programme under the Marie Sk{\l}odowska-Curie grant agreement No. 847471 (QUSTEC). 

\bmsection{Acknowledgments}
The devices were fabricated at the Binnig and Rohrer Nanotechnology Center (BRNC) at IBM Research Europe, Zurich. We thank the cleanroom operations team of the BRNC for their help and support.  We thank Michael Stiefel for making the FIB cross-section, Guanhao Huang for his support with the thermal response simulations, and Deividas Sabonis for critical reading of the manuscript. We are grateful to Clarissa Convertino, Felix Eltes, Tobias J. Kippenberg, and Junyin Zhang for helpful discussions. 

\bmsection{Disclosures}
The authors declare that there are no conflicts of interest related to this article.

\bmsection{Data availability}
Data supporting the plots within this paper and other findings of this study are available through Zenodo at  https://doi.org/10.5281/zenodo.14606207. Further information is available from the corresponding author upon reasonable request.

\bmsection{Credits}
This article may be downloaded for personal use only. Any other use requires prior permission of the author and AIP Publishing. This article appeared in Riedhauser, Möhl, Schading et al.; Absorption loss and Kerr nonlinearity in barium titanate waveguides; APL Photonics 1 January 2025; 10 (1): 016121 and may be found at https://doi.org/10.1063/5.0228990.

\end{backmatter}


\bibliography{sample}

\end{document}


\maketitle

\section{Derivation of expressions for resonator transmission and energy density}

In this section, we derive an expression for the energy density $\rho$ in the resonator, which is needed to calculate the Kerr nonlinear index $\overline{n_2}$ and the absorption rate $\kappa_\mathrm{abs}$ from the modulation-transfer data. 
In doing so, we use input-output theory to obtain an expression for the resonator transmission function that is used to fit the resonances and extract the cavity parameters that enter into the expression for $\rho$.

\begin{figure}[htbp] 
	\centering
	\includegraphics[width=0.4\textwidth]{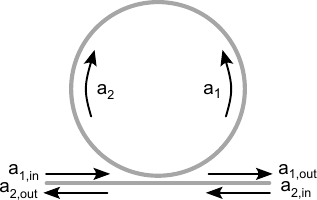} 
	\caption{ \centering Schematic of the forward and backward propagating modes.}
	\label{fig:input-output-scheme}
\end{figure}

We consider a single resonance of the ring resonator spectrum (Fig. 1d in the main text). As the pump and probe lasers are tuned to different resonances in the experiment, the treatment applies to either of the two.
The steady state equations of motion for the amplitudes of the forward and backward propagating resonator fields, $a_1$ and $a_2$, respectively, are given by  
\begin{equation}
    \label{eq:eq_motion}
    \renewcommand{\arraystretch}{2} 
    \begin{array}{ll}
        0 = -\dfrac{\kappa}{2} a_1 + i\Delta_1 a_1 + \sqrt{\kappa_{\mathrm{ex}}} a_{\mathrm{1,in}} + i \mu a_2 &  \\
        0 = -\dfrac{\kappa}{2} a_2 + i\Delta_2 a_2  + \sqrt{\kappa_{\mathrm{ex}}} a_{\mathrm{2,in}} + i \mu a_1,    & 
       \end{array}
\end{equation}
where $\kappa = \kappa_{\mathrm{ex}} + \kappa_0$ is the total cavity decay rate, $\kappa_{\mathrm{ex}}$ is the external coupling rate of the bus to the ring, and $\kappa_0$ is the internal loss rate. We assume that the forward and backward propagating modes have the same values of $\kappa_0$ and $\kappa_{\mathrm{ex}}$.  The detuning of the laser with respect to the cavity modes $a_1$ and $a_2$ are $\Delta_1$ and $\Delta_2$, respectively.  The coupling rate between forward and backward propagating fields in the resonator is denoted as $\mu$, and the respective input field amplitudes are $a_{\mathrm{1,in}}$ and $a_{\mathrm{2,in}}$. 
The output field amplitudes are related to the inputs by the following input-output relations:
\begin{equation}
    \label{eq:input_output}
    \renewcommand{\arraystretch}{2} 
    \begin{array}{ll}
        a_{\mathrm{1,out}} = a_{\mathrm{1,in}} - \sqrt{\kappa_{\mathrm{ex}}} a_1 &  \\
        a_{\mathrm{2,out}} = a_{\mathrm{2,in}} - \sqrt{\kappa_{\mathrm{ex}}} a_2.    & 
       \end{array}
\end{equation}
Figure \ref{fig:input-output-scheme} illustrates these modes for our ring resonator. 

In the experiment, the probe and pump lasers are counterpropagating, so the pump laser couples to the resonator via $a_{\mathrm{1,in}}$ and the probe laser via $a_{\mathrm{2,in}}$.
If we consider the pump resonance, for example, and set $a_{\mathrm{2,in}}=0$,  
we find the following expression for $a_1$ and $a_2$ as a function of $a_{\mathrm{1,in}}$:

\begin{eqnarray}
     a_1 &=& \dfrac{\sqrt{\kappa_{ex}}}{\dfrac{\kappa}{2} - i\Delta_1 + \dfrac{\mu^2}{\dfrac{\kappa}{2} - i\Delta_2}} a_{1,\mathrm{in}}\\
     a_2 &=& \dfrac{i\mu}{\dfrac{\kappa}{2} - i\Delta_2} a_1.
\end{eqnarray}
Experimentaly, the input and output fibers are edge coupled to the chip, and we observe that the overall fiber-to-chip-to-fiber transmission (-20 dB) varies from device to device by less than $\pm1$ dB.  We therefore assume equal fiber-to-chip coupling loss on both sides of the chip, so that 
$a_{\mathrm{1,in}}=\sqrt{P_{\mathrm{bus}}}$, 
where $P_{\mathrm{bus}}= \sqrt{P_{\mathrm{in}}P_{\mathrm{out}}}$ is the power in the bus waveguide, and $P_{\mathrm{in}}$ and $P_{\mathrm{out}}$ are the powers in the incoming and outgoing fibers, respectively.
We in turn estimate the error in the energy density in the device to also be $\pm1$ dB.

\begin{figure}[h] 
    \centering
    \includegraphics[width=0.7\textwidth]{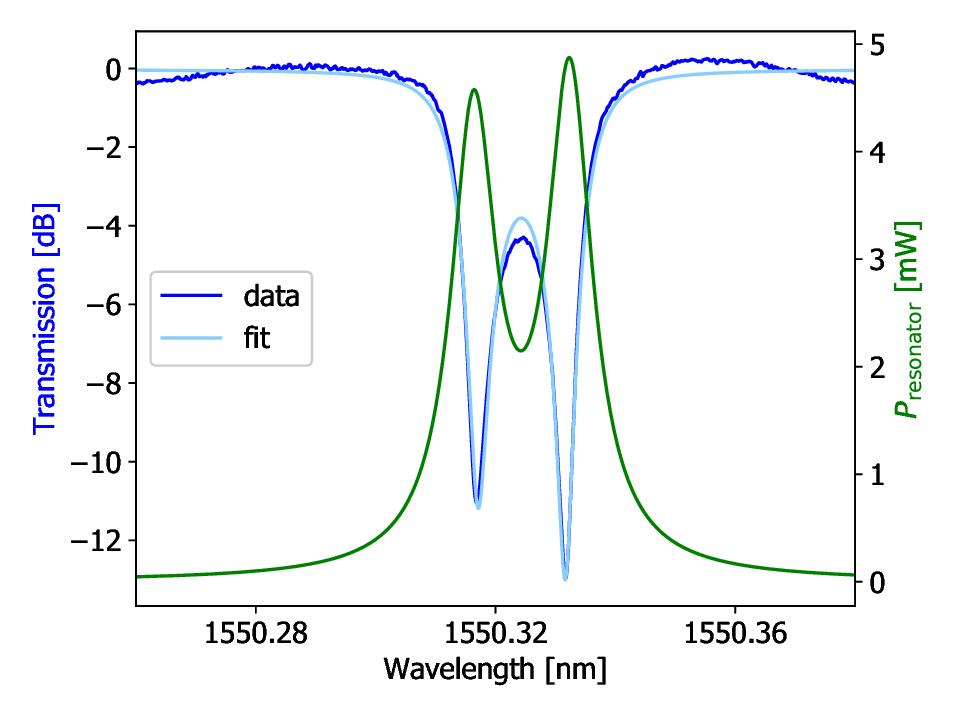} 
    \caption{ Measured transmission data (dark blue) and fit (light blue) of a typical resonance of a ring resonator. The trace in green depicting the circulating optical power in the resonator has pronounced peaks at the transmission minima. The parameters obtained from the fit are: $\kappa_{\mathrm{ex}} = 822 \pm 1$ MHz, $\kappa_0 = 314 \pm 1$ MHz and $\mu = 998 \pm 2$ MHz. The bus power is $P_{\mathrm{bus}}= 0.10 \pm 0.03$~mW. }
    \label{fig:S_fit_power}
\end{figure}

The transmission of the forward propagating field can be expressed as:
\begin{equation}
    T = \snorm{\frac{a_{\mathrm{1,out}}}{a_{\mathrm{a,in}}}}^2 = \snorm{1-\dfrac{\kappa_{\mathrm{ex}}}{\dfrac{\kappa}{2} - i\Delta_1 + \dfrac{\mu^2}{\dfrac{\kappa}{2} - i\Delta_2 }}}^2
    \label{transmission}
\end{equation}
A typical transmission spectrum together with a fit using Eq.~\ref{transmission} is shown in Fig.~\ref{fig:S_fit_power}. For each modulation-transfer experiment, a pump transmission spectrum was measured and the cavity parameters $\kappa_\mathrm{ex}$, $\kappa_\mathrm{0}$, and $\mu$ were determined.

The circulating power inside the resonator was computed using
\begin{equation}
P_{\mathrm{resonator}} = \dfrac{\snorm{a_1}^2 + \snorm{a_2}^2}{\tau_{rt}},
\label{Pres}
\end{equation}
where the cavity round-trip time is $\tau_{\mathrm{rt}} = \dfrac{n_{\mathrm{g,eff}}L}{c}$. Here, $n_{\mathrm{g,eff}}$ is the effective group index of the mode, $L$ is the length of the resonator, and $c$ is the speed of light.
The effective group index is given by
\begin{equation}
n_{\mathrm{g,eff}} = n_{\mathrm{eff}} - \lambda \dfrac{\mathrm{d}n_{\mathrm{eff}}}{\mathrm{d}\lambda}.
\label{eq:effective_group_index}
\end{equation}
Using COMSOL Multiphysics\textsuperscript{\textregistered} finite element simulations, we obtained $n_{\mathrm{eff}}= 1.67$ and $n_{\mathrm{g,eff}}= 2.19$ at 1550\,nm for our waveguide geometry, for which geometric dispersion dominates material dispersion by more than a factor of 10.

The expressions for $\overline{n_2}$ and $\kappa_\mathrm{abs}$ depend on the energy density
\begin{equation}
    \rho = \dfrac{\snorm{a_1}^2 + \snorm{a_2}^2}{V_{\mathrm{eff}}},
\end{equation}
where $V_\mathrm{eff}$ is the effective mode volume. To compute $V_\mathrm{eff}$, we write the electric fields $\mathbf{E}_j = a_j \mathbf{F_j}(r, \theta, z) e^{-i\omega t}$ of modes $j=1,2$ with spatial distributions $\mathbf{F_j}(r, \theta, z)$ in cylindrical coordinates \cite{Gao:1}. Since we work in the traveling-wave basis, $\snorm{\mathbf{F_1}(r,\theta,z)}^2$ = $\snorm{\mathbf{F_2}(r,\theta,z)}^2 \equiv \snorm{\mathbf{F}(r,z)}^2$, independent of $\theta$. $\mathbf{F}$ is normalized such that $\dfrac{\epsilon_0}{2} \int n_0^2 \snorm{\mathbf{F}}^2\mathrm{d}V = 1$. Simulating $\mathbf{F}$ in a radial cross section using COMSOL Multiphysics\textsuperscript{\textregistered}, we evaluate
\begin{equation}
V_{\mathrm{eff}} = \dfrac{\int n_0^2 \snorm{\mathbf{F}}^2 \mathrm{d}V \int \snorm{\mathbf{F}}^2 \mathrm{d}V}{\int n_0^2 \snorm{\mathbf{F}}^4\mathrm{d}V},
\label{Veff}
\end{equation}
where the volume integrals extends over the entire resonator, and $n_0$ is the local material refractive index. We take $n_{0,\mathrm{BaTiO_3}}$ = 2.28 \cite{eltes_integrated_2020} and $n_{0,\mathrm{SiO_2}}$ = 1.44 \cite{Malitson:65}. \\

\section{Intrinsic loss rates for measured devices}
\begin{figure}[htbp] 
    \centering
    \includegraphics[width=1.0\textwidth]{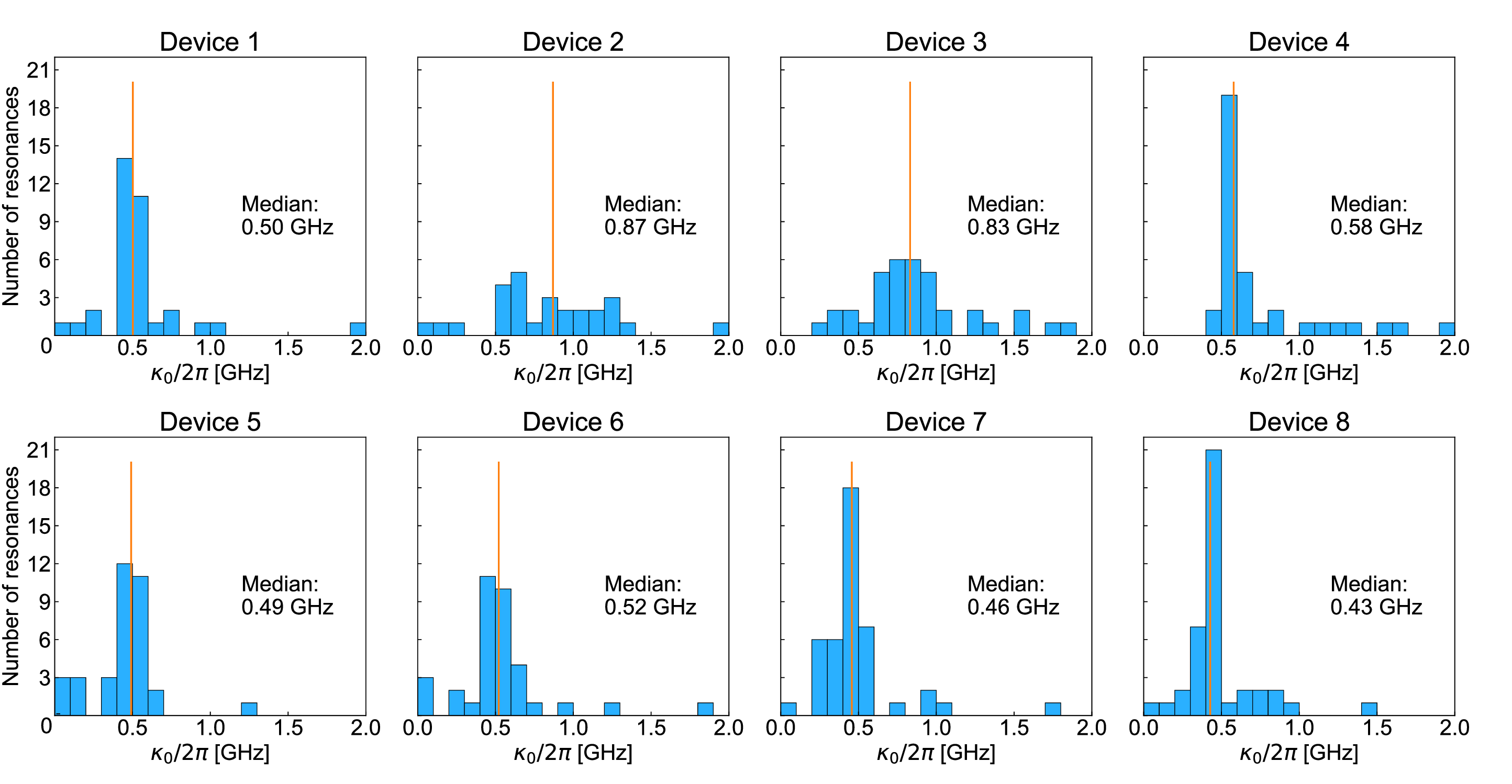} 
    \caption{ \centering Intrinsic loss rate distribution for all measured devices.   }
    \label{fig:hist_loss_rates}
\end{figure}

In this section, we provide the statistics of the intrinsic loss rates for all  measured devices (Fig. \ref{fig:hist_loss_rates}). For every device, 40 resonances in the wavelength range from 1500 to 1600 nm were fitted. Values of $\kappa_0/2\pi$ for which the fit did not converge were discarded. \\

\section{Measurement procedure and parameters}

The following experimental procedure was used for the determination of $n_2$ and $\kappa_\mathrm{abs}$:
\begin{enumerate}
    \item Calibrate probe phase-modulation depth $\beta_\mathrm{PM}$ using heterodyne detection.
    \item Calibrate pump intensity-modulation depth  $\beta_\mathrm{IM}$ using heterodyne detection.
    \item Measure pump transmission spectrum and tune pump on resonance, measuring fiber input and output powers, $P_\mathrm{in}$ and $P_\mathrm{out}$, respectively, on power meter.
    \item Measure probe transmission spectrum and tune probe to flank of resonance, measuring fiber input and output powers, $P_\mathrm{in}$ and $P_\mathrm{out}$, respectively, on power meter.
    \item Measure modulation-transfer curve using vector network analyzer.
    \item Measure power spectrum of cross-phase modulation and probe-laser phase modulation on real-time spectrum analyzer.
\end{enumerate}

Table \ref{tab:3001} gives a set of typical parameter values that were used for the modulation transfer experiments. \\

\begin{table}[htbp]
	\centering
	\begin{tabular}{cc}
		\toprule
	    	Parameter 	& Value \\
	    	\midrule
		$\lambda_{\mathrm{pump}}$ 						& $1550.7 \ $nm \\
		$\lambda_{\mathrm{probe}}$ 					& $1537.4 \ $nm \\
		$P_{\mathrm{bus, pump}}$ 		& $-8.76$ \ dBm $\triangleq 133~\mu$W \\
		$P_{\mathrm{bus, probe}}$ 		& $-19.54$ \ dBm $\triangleq 11.11~\mu$W \\
        $\rho$ &  53 J/m$^3$ \\
	    	$\beta_{\mathrm{IM}}$ 			& $0.158$ \\
	    	$\beta_{\mathrm{PM}}$ 			& $0.118$ \\
		$\overline{n_0 n_g}$	& $4.009$ \\
        $V_{\mathrm{eff}}$	& $3.14 \times 10^{-16}$ m$^3$ \\
        $\Gamma(\nu_{\mathrm{IM}})$ & $0.91$ \\
	    	\bottomrule
	\end{tabular}
	\caption{Typical parameters used for the modulation transfer experiment.}	
	\label{tab:3001}
\end{table}

\newpage
\section{Evidence of photorefractive effect in barium titanate}

\begin{figure}[htbp] 
    \centering
    \includegraphics[width=0.4\textwidth]{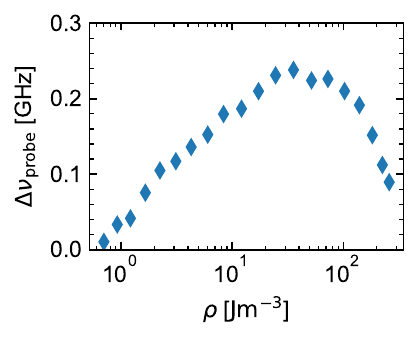} 
    \caption{ \centering  Probe resonance shift $\Delta \nu_{\mathrm{probe}}$ as a function of circulating energy density $\rho$ in the resonator.}
    \label{fig:photorefractive_power}
\end{figure}

Further evidence for the photorefractive effect in our platform is provided by a static pump-probe experiment in which we monitor the shift of the probe resonance frequency as a function of unmodulated pump power. Figure \ref{fig:photorefractive_power} shows the probe resonance shift with respect to the circulating energy density in the resonator. For low energy densities, we observe a blue shift with increasing energy density, which is consistent with the photorefractive effect \cite{Xu:21}. At larger energy densities, the photothermal effect dominates, leading to a red shift with increasing energy density. \\

\section{Calibration of the Kerr cross-phase cavity frequency modulation}
\label{sec:nu_XPM}
The modulation transfer response curve $A(\nu)$ is first corrected for the photoreceiver frequency response and then fitted using a three-pole model to determine the heights of the photothermal and Kerr plateaus, $A_\mathrm{PT}$ and $A_\mathrm{Kerr}$, respectively, and their ratio $\gamma = A_\mathrm{PT} / A_\mathrm{Kerr}$. The fit function reads
\begin{equation}
   A(\nu) = \left| \dfrac{A_\mathrm{PT}}{(1 + i \nu/\nu_1)^{c_1}(1 + i \nu/\nu_2)^{c_2}} + \dfrac{A_\mathrm{Kerr}}{1 + i \nu/\nu_3} \right|,
\end{equation}
where $\nu_1$ and $\nu_2$ are thermal cutoff frequencies and $\nu_3$ is the cutoff frequency corresponding to the cavity bandwidth. The exponents $c_1$  and $c_2$ have values near 0.5 and are also fitted to correctly describe the experimental data. The fit results are listed in Table \ref{tab:response_fit}

\begin{table}[htbp]
	\centering
	\begin{tabular}{cc}
		\toprule
		Parameter&Value\\
		\midrule
		$\gamma$&18.86\\
		$\nu_1$&11.35~kHz\\
		$\nu_2$&652.0~kHz\\
		$\nu_3$&434.2~MHz\\
		$c_1$&0.4057\\
		$c_2$&0.4922\\
		\bottomrule
	\end{tabular}
	\caption{Response function fit parameters.}	
	\label{tab:response_fit}
\end{table}

\newpage
The Kerr-induced cavity frequency modulation is calculated using \cite{Shams-Ansari}
\begin{equation}
\delta \nu_{\mathrm{XPM}} = \beta_{\mathrm{PM}} \nu_{\mathrm{PM}} \Gamma(\nu_{\mathrm{IM}}) \xi^{1/2},
\label{freq_calib}
\end{equation}
where $\xi = S(\nu_\mathrm{IM}) / S(\nu_\mathrm{PM})$ is the ratio of power spectral densities $S$ of the modulation-transfer signal and the reference phase-modulation signal measured on the real-time spectrum analyzer. The fraction of pure Kerr contribution to the total modulation-transfer signal $\Gamma(\nu_{\mathrm{IM}}) = A_\mathrm{Kerr} / A(\nu_\mathrm{IM})$ is retrieved from the fit to the measured response curve $A(\nu)$.  

To determine $\overline{n_2}$, the weighted index product $\overline{n_0n_g}$ appearing in Eq.~(1) of the main text must be known. It is calculated as \cite{Gao:1}
\begin{equation}
\overline{n_0n_g} = \dfrac{\int n_0n_g\snorm{\mathbf{F}}^2 \mathrm{d}V}{\int \snorm{\mathbf{F}}^2 \mathrm{d}V},
\label{eq:weighted_index_product}
\end{equation}
where $n_g$ is the group index and  $n_0$ is the linear refractive index at the resonance frequency.
Both indices are a spatially dependent functions given by the device geometry and materials. For the evaluation of the weighted index product, we take $n_{\mathrm{g,SiO_2}} = 1.46$ \cite{Malitson:65} and $n_{\mathrm{g,BaTiO_3}}$ = 2.33. The value for BaTiO$_3$ was determined by ellipsometry.\\

\section{Determination of the Kerr nonlinearity of barium titanate}
\begin{figure}[htbp] 
    \centering
    \includegraphics[width=0.6\textwidth]{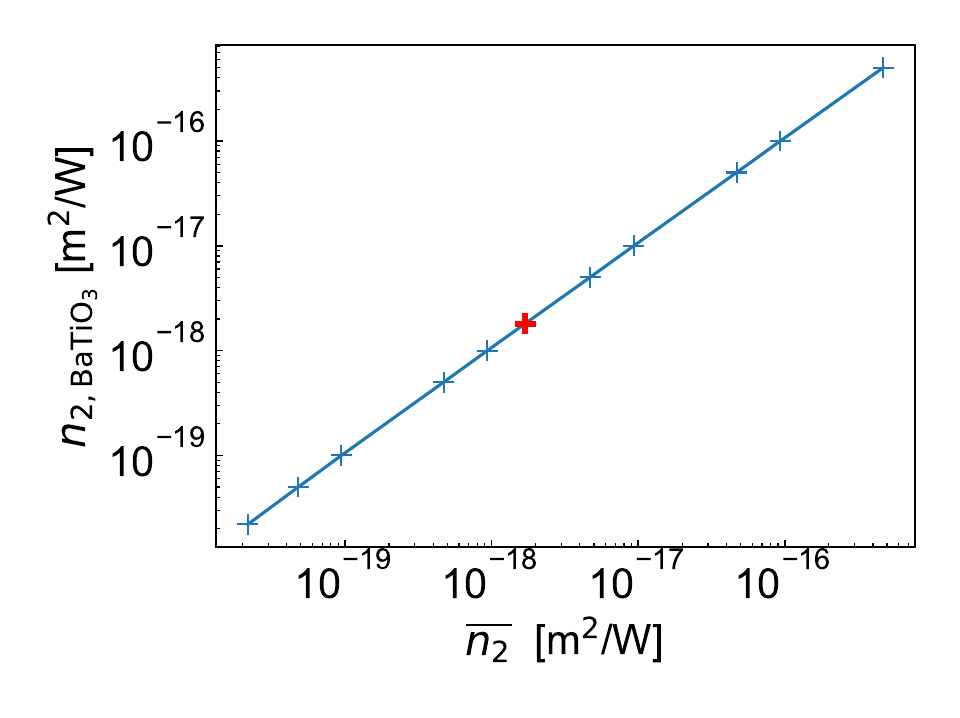} 
    \caption{Correspondence between the simulated weighted nonlinear index of the ridge waveguide $\overline{n_2}$ and the assumed Kerr nonlinear index $n_{\mathrm{2,BaTiO_3}}$ of BaTiO$_3$. Markers indicate the various values simulated.
    The line is a linear fit to the simulated data used for interpolation. The red marker indicates the values determined for $\overline{n_2}$ and $n_{\mathrm{2,BaTiO_3}}$.}
    \label{fig:n2_BTO}
    
\end{figure}

From the modulation transfer response curve, we determine the weighted nonlinear index $\overline{n_2}$ of the air-clad BaTiO$_3$-on-SiO$_2$ ridge-waveguide platform (see main text). We can extract the Kerr nonlinear index of BaTiO$_3$ using \cite{Gao:1}
\begin{eqnarray}
\overline{n_2} &=& \dfrac{\int n_0^2 n_2 \snorm{\mathbf{F}}^4 \mathrm{d}V}{\int n_0^2 \snorm{\mathbf{F}}^4 \mathrm{d}V} \label{eq:weighted_index_product}\\
&=& \dfrac{\int_\mathrm{BaTiO_3} n_{0,\mathrm{BaTiO_3}}^2 n_{2,\mathrm{BaTiO_3}} \snorm{\mathbf{F}}^4 \mathrm{d}V + \int_\mathrm{SiO_2} n_{0,\mathrm{SiO_2}}^2 n_{2,\mathrm{SiO_2}} \snorm{\mathbf{F}}^4 \mathrm{d}V}{\int_\mathrm{BaTiO_3} n_{0,\mathrm{BaTiO_3}}^2 \snorm{\mathbf{F}}^4 \mathrm{d}V + \int_\mathrm{SiO_2} n_{0,\mathrm{SiO_2}}^2 \snorm{\mathbf{F}}^4 \mathrm{d}V},
\end{eqnarray}
where $n_2$ is the spatially varying Kerr nonlinear index. Using COMSOL Multiphysics\textsuperscript{\textregistered} finite element simulations, we computed $\overline{n_2}$ for various values of $n_{\mathrm{2,BaTiO_3}}$ (Fig. \ref{fig:n2_BTO}), assuming $n_{\mathrm{2,SiO_2}}$ = $2.42 \cdot 10^{-20}$ m$^2$/W \cite{Flom:15}.  The simulated values were linearly interpolated to relate the measured $\overline{n_2}$ to $n_{\mathrm{2,BaTiO_3}}$.

\newpage
\section{Thermal analysis}
\subsection{Determination of the thermorefractive shift}
\begin{figure}[htbp] 
    \centering
    \includegraphics[width=0.6\textwidth]{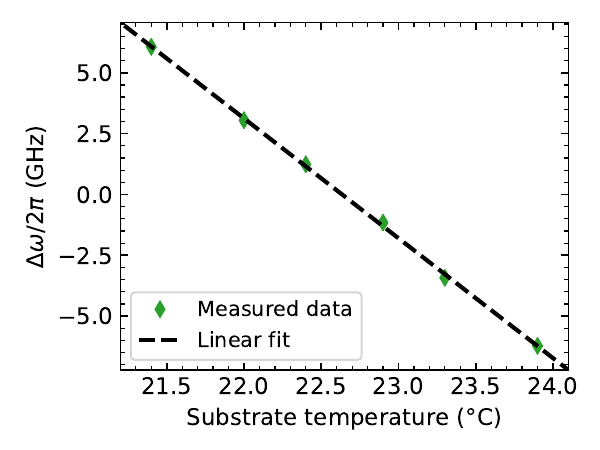} 
    \caption{ \centering Resonance frequency shift as function of device mounting plate temperature.}
    \label{fig:thermal_shift}
    
\end{figure}

In order to compute $\kappa_\mathrm{abs}$, the change in resonance frequency with temperature due to the thermorefractive effect $\dfrac{\mathrm{d}\nu}{\mathrm{d}T}$ is needed. We measure  $\dfrac{\mathrm{d}\nu}{\mathrm{d}T}$ by varying the temperature of the plate on which the device is mounted using a Peltier element and monitoring the resonance frequency shift at low optical intensity, for which there is no significant heating due to the laser light. (Fig. \ref{fig:thermal_shift}). 
BaTiO$_3$ waveguides are assumed to be in thermal equilibrium with the substrate after a waiting time of 5 minutes, as the resonance frequency shift remained stable in repeated measurements.

In contrast to the modulation transfer experiment, in which the absorbed light heats the device locally in the waveguide, here the silicon substrate temperature is also changing. This leads to an expansion of the substrate, which adds a contribution to the frequency shift due to an increase of the resonator length $L$. Therefore, the measured total thermally induced frequency dependence $\dfrac{\mathrm{d}\nu_{\mathrm{res}}}{\mathrm{d}T}$ is given by  
\begin{equation}
\dfrac{\mathrm{d}\nu_{\mathrm{res}}}{\mathrm{d}T} = \dfrac{\partial \nu_{\mathrm{res}}}{\partial n_{\mathrm{eff}}} \dfrac{\partial n_{\mathrm{eff}}}{\partial T}+ \dfrac{\partial \nu_{\mathrm{res}}}{\partial L} \dfrac{\partial L}{\partial T}, 
\end{equation}
where $  \dfrac{\partial \nu_{\mathrm{res}}}{\partial n_{\mathrm{eff}}} \dfrac{\partial n_{\mathrm{eff}}}{\partial T} = \dfrac{\mathrm{d}\nu}{\mathrm{d}T}$ is the 
rate of frequency shift with temperature due to the thermorefractive effect alone.
For the mode with azimuthal mode number $m$, the resonance frequency is given by $\nu_{\mathrm{res}} = \dfrac{c m}{L n_{\mathrm{eff}}}$. The coefficient of linear thermal expansion of silicon can be expressed as $\alpha_L = \dfrac{1}{L} \dfrac{\partial L}{\partial T}$. Using these expressions, it is straightforward to show that
\begin{equation}
\dfrac{\mathrm{d}\nu}{\mathrm{d}T} = \dfrac{\mathrm{d}\nu_{\mathrm{res}}}{\mathrm{d}T} + \alpha_L\nu_{\mathrm{res}}.
\label{eq:thermo}
\end{equation}
In other words, $\dfrac{\mathrm{d}\nu}{\mathrm{d}T}$
can be determined by measuring the total resonance shift and correcting it for substrate expansion. 
Figure \ref{fig:thermal_shift} shows the total resonance frequency shift as function of device mounting plate temperature. The slope gives $\dfrac{\mathrm{d}\nu_{\mathrm{res}}}{\mathrm{d}T} = -4.92$~GHz~K$^{-1}$. Using $\alpha_L = 2.6 \cdot 10^{-6}$~K$^{-1}$~\cite{alpha_l}, we find $\dfrac{\mathrm{d}\nu}{\mathrm{d}T} = -4.42$~GHz~K$^{-1}$.

\subsection{Simulation of the thermal response due to absorbed light}
\begin{figure}[htbp] 
    \centering
    \includegraphics[width=0.6\textwidth]{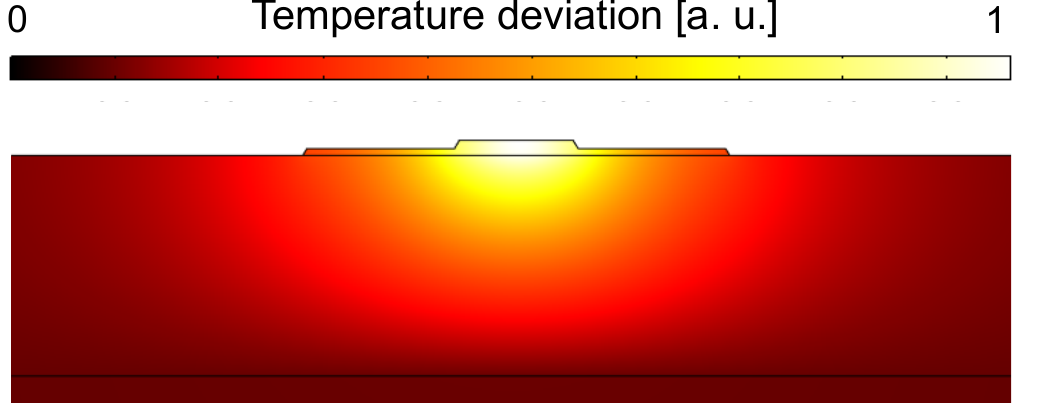} 
    \caption{ Normalized temperature deviation distribution for $\Tilde{T}(\Omega = 0, \mathbf{r})$ resulting from the absorbed heat $\delta P_{\mathrm{abs}}$.}
    \label{fig:thermal_simulations}
\end{figure}

To determine the rate of temperature change with absorbed optical power, $\dfrac{\mathrm{d}T}{\mathrm{d}P_{\mathrm{abs}}}$, used in Eq. 2 in the main text, we perform finite-element-method simulations.
We first simulate the electric field distribution  $\mathbf{F(\mathbf{r})}$. The spatially dependent absorped power given by
\begin{equation}
\delta P(\mathbf{r}) = \delta P_{\mathrm{abs}} \epsilon_0 \mathbf{F}(\mathbf{r}) \times \hat{\epsilon}\mathbf{F}(\mathbf{r})/2
\end{equation}
is then considered as the heat source \cite{PRXQuantum.3.020309}, with $\hat{\epsilon} = \hat{\epsilon_0} + \hat{\epsilon_1}$ being the permittivity.  The Fourier-domain heat equation is
\begin{equation}
i \Omega \rho_m C \Tilde{T} + k \Delta \Tilde{T} = \delta\Tilde{P},
\end{equation}
where $\rho_m$ is the material density, $C$ the heat capacity, $k$ the thermal conductivity, and $\Omega$ the modulation frequency.
Following the approach of refs. \cite{Gao:1, liu_high-yield_2021}, we use the Heat Transfer Module in COMSOL Multiphysics\textsuperscript{\textregistered} to solve this equation with the parameters in Table \ref{tab:material parameters} and obtain the temperature distribution $\Tilde{T}(\Omega,\mathbf{r})$ for a given absorbed power 
$\delta P_{\mathrm{abs}}$ (Fig.~\ref{fig:thermal_simulations}). 
We calculate the average temperature change as
\begin{equation}
    \delta T = \dfrac{\int n_0^2 \alpha_n \Tilde{T}(\Omega=0,\mathbf{r}) \snorm{\mathbf{F}}^2 \mathrm{d}V}{\int n_0^2 \alpha_n \snorm{\mathbf{F}}^2 \mathrm{d}V},
\end{equation}
where $\alpha_n = \dfrac{1}{n_0} \dfrac{\partial n_0}{\partial T}$ is the thermorefractive coefficient, as defined in \cite{Gao:1}. We find $\dfrac{\delta T}{\delta P_{\mathrm{abs}}} \approx \dfrac{\mathrm{d}T}{\mathrm{d}P_{\mathrm{abs}}} = 903$ K/W for our platform.

\begin{table}[h!]
    \centering
    \begin{tabular}{cccc}
    \toprule
     Material & Density (kg m$^{-3}$) & Specific heat capacity (J kg$^{-1}$ K$^{-1}$) & Thermal conductivity (W m$^{-1}$ K$^{-1}$) \\
     \midrule
     BaTiO$_3$ \cite{Strukov_2003} & 6020 & 450.27 & 5.1\\
     SiO$_2$ \cite{Gao:1} & 2200 & 740 & 1.4\\
     Si \cite{Gao:1} & 2330 & 700 & 130 \\
     \bottomrule
    \end{tabular}
	\caption{Material parameters used for simulation of the thermal response due to absorbed light.}	
	\label{tab:material parameters}
\end{table}

\newpage
\section{Extracted values of Kerr nonlinearity and absorption losses}
In Table \ref{tab:3002}, we give the weighted nonlinear index $\overline{n_2}$ and the nonlinear index of barium titanate $n_{\mathrm{2,BaTiO_3}}$ obtained for eight different devices.
Table \ref{tab:3004} lists the absorption loss rate $\kappa_{\mathrm{abs}}$, the intrinsic loss rate  $\kappa_0$, the absorption-limited propagation loss 
\begin{equation}
	\alpha_{\mathrm{abs}} = 10 \cdot \mathrm{log_{10}(e)} \cdot \dfrac{n_{\mathrm{g,eff}}}{c} \kappa_{\mathrm{abs}},
\end{equation}
and the ratio of the photothermal and Kerr plateaus $\gamma$ for the same devices. 
The errors in $\overline{n_2}$, $n_{\mathrm{2,BaTiO_3}}$, $\kappa_{\mathrm{abs}}/2\pi$, and $\alpha_{\mathrm{abs}}$ are primarily influenced by the uncertainties in energy densities. The reported errors reflect the propagation of a 1-dB error in energy density. 
The spread in intrinsic and absorption-limited quality factors across the devices may be attributed to some extent to process non-uniformities over the 1~cm$\times$1~cm substrates resulting from chip-level fabrication.
 
\begin{table}[htbp]
	\centering
	\begin{tabular}{cccc}
		\toprule
	    	Device no. 	&  $\overline{n_2}$  (10$^{-18}$ m$^2$ W$^{-1}$) &  $n_{\mathrm{2,BaTiO_3}}$ (10$^{-18}$ m$^2$ W$^{-1}$)\\
	    	\midrule
	    	$1$ 			& $1.3 \pm 0.3$ & $1.4 \pm 0.3$ \\
	    	$2$ 			& $2.3 \pm 0.6$ & $2.4 \pm 0.6$ \\
	    	$3$ 			& $1.7 \pm 0.4$ & $1.8 \pm 0.4$ \\
	    	$4$ 			& $1.5 \pm 0.4$ & $1.6 \pm 0.4$ \\
	    	$5$ 			& $2.0 \pm 0.5$ & $2.1 \pm 0.5$\\
	    	$6$ 			& $1.7 \pm 0.4$ & $1.8 \pm 0.4$\\
	    	$7$ 			& $1.1 \pm 0.3$ & $1.2 \pm 0.3 $ \\
	    	$8$ 			& $2.0 \pm 0.5$ & $2.1 \pm 0.5$  \\
	    	\bottomrule
	\end{tabular}
	\caption{Extracted values for the weighted nonlinear index $\overline{n_2}$ and the nonlinear index of BaTiO$_3$ $n_{\mathrm{2,BaTiO_3}}$ for several devices.}	
	\label{tab:3002}
\end{table}

\begin{table}[htbp]
	\centering
	\begin{tabular}{ccccc}
		\toprule
	    	Device no. 	&  $\kappa_{\mathrm{abs}}/2\pi \ $(MHz)	& $\kappa_0/2\pi \ $(MHz)	& $\alpha_{\mathrm{abs}} \ (\frac{\mathrm{dB}}{\mathrm{m}})$ 	& $\gamma$ \\
	    	\midrule
	    	$1$ 				& $48 \pm 12$ 					& $480$ 				& $9.7 \pm 2.5$ 				& $7.6 $ \\
	    	$2$ 			& $53 \pm 14$ 					& $1060$ 				& $10.6 \pm 2.7$ 				& $8.4$ \\
	    	$3$ 			& $53 \pm 14$ 					& $662.5$ 				& $10.5 \pm 2.7$ 				& $8.4 $ \\
	    	$4$ 			& $97 \pm 25$				 	& $510$ 				& $19.3 \pm 5.0$ 				& $15.3$ \\
	    	$5$ 			& $53 \pm 14$ 					& $441$ 				& $10.5 \pm 2.7$ 				& $8.4$ \\
	    	$6$ 			& $139 \pm 36$ 					& $463$ 				& $27.6 \pm 7.1$ 				& $22.0$ \\
	    	$7$ 			& $70 \pm 18$ 					& $437$ 				& $14.0 \pm 3.6$ 				& $11.2$ \\
	    	$8$ 			& $56 \pm 15$ 					& $373$ 				& $11.1 \pm 2.9$ 				& $8.9$ \\
	    	\bottomrule
	\end{tabular}
	\caption{Extracted values for absorption rate  $\kappa_{\mathrm{abs}}$, intrinsic loss rate $\kappa_0$,  absorption-limited propagation loss $\alpha_{\mathrm{abs}}$, and the ratio $\gamma$ of the levels of the photothermal and Kerr response curve plateaus for several devices.}
	\label{tab:3004}
\end{table}

\newpage
\section{Comparison of literature values for the Kerr nonlinearity}

There are several factors that could account for the discrepency between the nonlinear index $n_{\mathrm{2,BaTiO_3}}$ we measure and the values reported in the literature (see Table \ref{tab:literature comparison}).  These include:
\begin{enumerate}
	\item Material dimensions: Ganeev  \emph{et al.} \cite{Ganeev:08} reported two different values for the nonlinearity, one for a bulk crystal and one for nanoparticles in a suspension. The value for the nanoparticles is more than 5000 times larger than for the bulk sample. Yust \emph{et al.} \cite {yust_enhancement_2012} also found that nanoparticles of BaTiO$_3$ can have giant nonlinearities. This suggests a possible dependence of the nonlinearity on the size of the structure being measured. Our system makes use of ridge waveguides with a ridge width of 1.6~$\mu$m, a total height of 225~nm, and a ridge height of 110~nm. In other words, the dimensions of the system are between those of a bulk crystal and nanoparticles, but closer to those of nanoparticles.
	\item  Material: The materials studied in the literature were synthesized with various methods, including solution chemistry, potentially resulting in significant differences in composition with respect to our material.
	\item Measurement method: All previous studies used the z-scan method, in some cases with a pulsed light source, in other cases with a continuous-wave (cw) source. Analysis of these measurements requires taking into account effects of pulse duration as well as thermo-optical nonlinearities associated with local heating. The modulation-transfer method that we use is fundamentally different.
	\item Wavelength: The nonlinear index of a material generally depends on wavelength. The dependence can be significant, particularly in the vicinity of an absorption feature of the material.
\end{enumerate}
Given the numerous variables to consider, a strict comparison of our results to previous literature values is not possible.
\begin{table}[h!]
	\begin{hyphenrules}{nohyphenation}
	\makebox[\linewidth][c]{
	\begin{tabular}{lp{6.1cm}>{\centering}p{2cm}cc}
		\toprule
		Work & Sample & Method & Wavelength & $n_{\mathrm{2,BaTiO_3}}$\\
		& & & (nm) & (10$^{-20}$ m$^2$ W$^{-1}$)\\
		\midrule
		This work & Ridge waveguide (ridge width = 1.6~$\mu$m, total height = 225~nm, ridge height = 110~nm) & Modulation transfer & 1550 & 180\\
		Zhang \emph{et al.} \cite{zhang_nonlinear_2000} & Thin film (thickness = 530~nm) & z-scan, pulsed & 1064 & Not observed\\
		Ganeev \emph{et al.} \cite{Ganeev:08} & Bulk BaTiO$_3$ (thickness = 2~mm) & z-scan, pulsed & 790 & 6\\
		Ganeev \emph{et al.} \cite{Ganeev:08} & Suspension of BaTiO$_3$ nanoparticles in ethylene glycol (mean diameter = 92 nm, diameter range  = 50-160 nm, concentration = 5~mM)& z-scan, pulsed & 790 & $^{a)}$ 3.4 $\times$ 10$^4$\\
		Yust \emph{et al.} \cite{yust_enhancement_2012} & Suspension of BaTiO$_3$ nanoparticles in water (mean diameter =  500~nm, concentration not reported) & z-scan, cw & 532 & 4.7 $\times$ 10$^6$\\
		Yust \emph{et al.} \cite{yust_enhancement_2012} & Suspension of BaTiO$_3$ nanoparticles in water (mean diameter = 200~nm, concentration not reported) & z-scan, cw & 532 & 6.6 $\times$ 10$^6$\\
		Yust \emph{et al.} \cite{yust_enhancement_2012} & Suspension of BaTiO$_3$ nanorods in water (diameter = 50 nm, mean length = 200~nm, concentration not reported) & z-scan, cw & 532 & 2.5 $\times$ 10$^6$\\
		\bottomrule
		\multicolumn{5}{l}{a) Assumes 10$^{-3}$ part by volume of BaTiO$_3$ as reported in ref. \cite{Ganeev:08}.}
		\end{tabular}
	}
\end{hyphenrules}
	\caption{Compilation of literature values of $n_{\mathrm{2,BaTiO_3}}$.}	
	\label{tab:literature comparison}
\end{table}

\newpage
\bibliography{sample}